\documentclass[%
 reprint,
 amsmath,amssymb,
 aps,
]{revtex4-1}

\usepackage{graphicx}
\usepackage{dcolumn}
\usepackage{bm}
\usepackage{physics}
\usepackage{amsmath}

\begin{document}


\title{Cavity-induced artificial gauge field in a Bose-Hubbard ladder}

\author{Catalin-Mihai Halati}
\author{Ameneh Sheikhan}
 \author{Corinna Kollath}
\affiliation{%
 HISKP, University of Bonn, Nussallee 14-16, 53115 Bonn, Germany
}%

\date{\today}

\begin{abstract}
We consider theoretically ultracold interacting bosonic atoms confined to quasi-one-dimensional ladder structures formed by optical lattices and coupled to the field of an optical cavity. The atoms can collect a spatial phase imprint during a cavity-assisted tunneling along a rung via Raman transitions employing a cavity mode and a transverse running wave pump beam. By adiabatic elimination of the cavity field we obtain an effective Hamiltonian for the bosonic atoms, with a self-consistency condition. Using the numerical density matrix renormalization group method, we obtain a rich steady state diagram of self-organized steady states. Transitions between superfluid to Mott-insulating states occur, on top of which we can have Meissner, vortex liquid, and vortex lattice phases. Also a state that explicitly breaks the symmetry between the two legs of the ladder, namely the biased-ladder phase is dynamically stabilized. We investigate the influence that a trapping potential has on the stability of the self-organized phases. 

\end{abstract}

\pacs{Valid PACS appear here}
\maketitle


\section{\label{sec:Introduction}Introduction}
Ultracold atoms coupled to an optical cavity mode have proved to be an exciting field of physics \cite{RitschEsslinger2013}. In recent experiments \cite{ BaumannEsslinger2010, KlinderHemmerich2015}, ultracold bosonic atoms placed in an optical cavity have 
realized a Dicke phase transition \cite{DomokosRitsch2002, DimerCarmichael2007, NagyDomokos2008, PiazzaZwerger2013}. By additionally confining the atomic gas with external optical lattice potentials, a modified Bose-Hubbard model with long-range interactions has been experimentally realized \cite{KlinderHemmerichPRL2015, LandigEsslinger2016}. This has been investigated theoretically \cite{ElliotMekhov2016, BakhtiariThorwart2015} and, in particular, the effect of the long-range interactions on the superfluid to Mott insulator transition has been analyzed \cite{MaschlerRitsch2005, MaschlerRitsch2008, LarsonLewenstein2008, NiedenzuRitsch2010, SilverSimons2010, VidalMorigi2010, LiHofstetter2013}. More complex combined cavity-atom systems have been proposed, as the organization of bosonic atoms in triangular or hexagonal lattices \cite{SafaeiGremaud2015, LeonardDonner2017, LeonardEsslinger2017}, or of fermionic atoms into superradiant phases \cite{LarsonLewensteinPRA2008, MullerSachdev2012, KeelingSimons2014, PiazzaStrack2014, ChenZhai2014}. Disordered structures might be realized in setups such as multimode cavities \cite{Gopalakishnan2009, NimmrichterArndt2010, StrackSachdev2011, Gopalakishnan2011, HabibianMorigi2013, JanotRosenow2013, BuchholdDiehl2013}. Phases for which the spin-orbit coupling plays an important role have been discussed for standing-wave cavities \cite{DengYi2014, DongPu2014, PanGuo2015, PadhiGhosh2014}, or ring cavities \cite{MivehvarFeder2014, MivehvarFeder2015}. 

Another class of systems which has attracted a lot of interest recently, are ultracold atomic systems coupled to artificial gauge fields \cite{DalibardOhberg2011, GoldmanSpielman2014}. The generation of an artificial gauge field has been realized in different ways such as Raman coupling \cite{LinSpielman2009, LinSpielmanNature2009}, lattice shaking \cite{StruckSengstock2011}, or laser-assisted hopping \cite{AidelsburgerBloch2011,MiyakeKetterle2013}. These artificial gauge fields for neutral atoms have similar effects as magnetic fields for charged particles. 

The minimal geometry in which the effects of gauge fields are important is the quasi-one-dimensional ladder structure. For weakly interacting ultracold bosons on a ladder a Meissner and a vortex superfluid phase were observed \cite{AtalaBloch2014}. Theoretically, other exciting phases such as vortex lattice and biased ladder superfluid phase, Meissner and vortex Mott insulator were predicted to occur \cite{OrignacGiamarchi2001, TokunoGeorges2014, PiraudSchollwock2015, GreschnerVekua2015, GreschnerVekua2016, Dhar2012, Dhar2013, WeiMuller2014, UchinoTokuno2015, Uchino2016, PetrescuHur2015, StrinatiMazza2016, PetrescuHur2017, DioChiofalo2015, OrignacChiofalo2016, OrignacPalo2017}.

In recent years proposals have been put forward for the dynamic generation of gauge fields by a cavity-assisted tunneling. 
The artificial magnetic field emerges dynamically due to the feedback mechanism between the cavity field and the motion of atoms \cite{CorinnaAmenehStefanPRL2016, KollathBrennecke2016,WolffKollath2016, SheikhanKollath2016, ZhengCooper2016, BallantineKeeling2017}.  
The steady state diagram and the dynamics has been determined in the case of noninteracting fermions  including states with chiral currents on a ladder geometry  \cite{CorinnaAmenehStefanPRL2016, KollathBrennecke2016,WolffKollath2016} or non-trivial topological properties in two dimensions \cite{SheikhanKollath2016}. 

In the present work we consider interacting bosons on a ladder structure coupled to a cavity mode and explore the steady state diagram for different interaction strengths, different magnetic fluxes and different fillings. We characterize the self-organized phases that arise and we investigate the stability of these phase in the coupled atomic cavity system. Additionally, we investigate the influence a trapping potential has on the stability of the phases, since in most present day setups such trapping potentials are present. 

The structure of the paper is as follows, in Sec.~\ref{2a} we describe the setup of the bosonic atoms in the optical cavity and the theoretical model. In Sec.~\ref{2b} we derive an effective model for the atomic degrees of freedom by performing the adiabatic elimination of the cavity field. In Sec.~\ref{2c} we derive a stability condition for the steady states. In Sec.~\ref{sec:selfconst} we show how one can relate the solutions of the effective model to the steady states of the coupled system and in Sec.~\ref{sec:effectiveH} we discuss the properties of the effective Hamiltonian. In Sec.~\ref{sec:method} we give typical parameters used within the numerical density matrix renormalization group method. The stable self-organized phases with a finite cavity field and their properties are presented in Sec.~\ref{sec:results} and the influence of a trapping potential is discussed. 

\section{\label{sec:Model}Model and Method}

\subsection{\label{sec:setup}Description of the setup}
\label{2a}

\begin{figure}[hbtp]
\centering
\includegraphics[width=.5\textwidth]{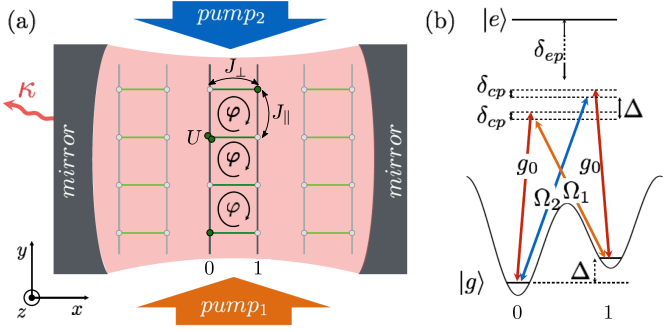}
\caption{(a) Sketch of the setup. 
The bosonic atoms in an optical cavity are placed in an optical super-lattice potential which creates an array of ladders. The atoms tunnel along the legs with the amplitude $J_\|$, along the rungs via the cavity-induced tunneling with an effective amplitude $J_\perp$ and have an on-site interaction of strength $U$. (b) Level scheme of the cavity-induced tunneling: $\ket{g}$, $\ket{e}$ denote the ground and excited internal electronic states. The energy offset between two neighboring wells, $\Delta$, strongly suppresses the tunneling along the rungs. This is restored by two Raman processes each of which involve the cavity mode with vacuum Rabi frequency $g_0$ and a transverse pump beam with Rabi frequency $\Omega_{1,2}$, respectively. $\delta_{ep}$ and $\delta_{cp}$ are the frequencies of the excited state and the cavity mode in the rotating frame. 
 }
\label{fig:setup}
\end{figure}

We study an ultracold bosonic gas placed in an optical cavity and additionally subjected to optical lattice potentials (Fig.~\ref{fig:setup}). A similar setup has been introduced and discussed in previous work in the context of fermionic atoms  \cite{CorinnaAmenehStefanPRL2016, KollathBrennecke2016,WolffKollath2016}. The optical super-lattice potentials confine the atoms to an array of decoupled ladders. The first step in obtaining this structure is to create two-dimensional decoupled layers by applying a strong optical lattice potential along the $z$-direction. One applies an optical lattice of wavelength $\lambda_y$ along the $y$-direction. A superposition of two optical lattices with wavelengths $\lambda_x$ and $\lambda_x/2$ is applied along the $x$-direction, such that decoupled double wells are formed with an energy offset $\Delta$ between the two wells. The lattice height along the $y$-direction is sufficiently low such that the atoms tunnel with amplitude $J_\|$ between neighboring sites. The potential offset between the two wells along $x$-direction strongly suppresses the tunneling along the rungs. The tunneling is restored by two balanced Raman transitions each of them involving a standing-wave cavity mode and a running-wave pump laser (Fig.~\ref{fig:setup}(b)). The cavity mode has the frequency $\omega_c$, vacuum Rabi frequency $g_0$ and the wave-vector ${\bf k}_c=k_c{\bf e}_x$ along the $x$-direction, where ${\bf e}_i$, with $i=x,y,z$, denote the unit vectors along the three spatial directions. All other cavity modes are assumed to be far detuned from the possible transitions and are not considered. The pump laser beams have the frequency $\omega_{p,i=1,2}$ and the wave-vector ${\bf k}_{p,i}=k_{p,i}{\bf e}_y$, with $i=1,2$. The pump and cavity modes are considered to be far detuned from the internal atomic transition, i.e. $\omega_e\gg\omega_c,\omega_{p,i=1,2}$, thus the excited state population is negligible. The detuning between the cavity mode and the first pumping beam is chosen such that it is close to the potential offset, $\hbar(\omega_c-\omega_{p,1})\approx-\Delta$ and the difference between the two pump beam frequencies is $\hbar(\omega_{p,2}-\omega_{p,1})\approx 2\Delta$. Let us note that the offset $\Delta$ needs to be choosen large enough in order to assure a resonant tunneling process even in the presence of the finite cavity line width. 

A cavity-induced Raman tunneling along the rungs of the ladder is obtained, via the feedback mechanism between the motion of the atoms and the cavity mode. In the following we will use the rotating frame with the frequency $\omega_p=(\omega_{p,2}+\omega_{p,1})/2$ and $\lambda_{p}= \lambda_{p,1,(2)}$ as the wavelengths of the pump beams are approximately the same, i.e.~$\lambda_{p,1}\approx\lambda_{p,2}$ \cite{notewavelength}.

During the Raman processes a spatially dependent phase factor $\text{e}^{-i \Delta\bf{k\cdot r}}$ is imprinted onto the atomic wave-function, where $\Delta{\bf k}=\pm k_c{\bf e}_x+k_{p}{\bf e}_y$. The cavity mode does not give a contribution if the tunneling around a plaquette is considered. However, due to the running-wave nature of the pump beam the atoms collect a phase $\varphi(j+1)=\pi \lambda_y/\lambda_p(j+1) $ tunneling on the rung $j+1$ and a phase $-\varphi j$ on the rung $j$, where $\varphi\simeq\pi \lambda_y/\lambda_p$,  The accumulated phase $\varphi$ is equivalent to the Aharonov-Bohm phase for charged particles subjected to a magnetic field. Thus, the bosonic atoms experience an artificial magnetic field in the presence of a finite cavity field. In an experimental realization, the flux $\varphi$ can be varied by modifying the angle of the pump beams with respect to the $x$-$y$-plane.

We perform an expansion in the Wannier basis of the atomic field operators, obtaining a model for the atomic-cavity system in the tight-binding description. The Hamiltonian describing the coupled system reads

\begin{align}
\label{eq:Hamiltonian}
&H=H_c+H_\parallel+H_{int}+H_{trap}+H_{ac} \\
&H_c= \hbar\delta_{cp} a^\dagger a\nonumber\\
&H_\parallel=-J_\parallel \sum_{j,m=0,1} (b_{m,j}^\dagger b_{m,j+1} + b_{m,j+1}^\dagger b_{m,j})\nonumber\\
&H_{int}=\frac{U}{2} \sum_{j,m=0,1} n_{m,j}(n_{m,j}-1)\nonumber\\
&H_{trap}=\frac{4 V_{trap}}{L^2}\sum_{j,m=0,1}\qty(j-j_0)^2 n_{m,j}\nonumber\\
&H_{ac}=  -\hbar\tilde{\Omega} ( a + a^\dagger) ( K_\perp + K_\perp^\dagger)\nonumber\\
&K_\perp=  \sum_{j} e^{i\varphi j}b_{0,j}^\dagger b_{1,j}\nonumber.
\end{align}

The bosonic operators $a$ and $a^\dagger$ are the annihilation and creation operators for the cavity photon mode. The term $H_c$ gives the dynamics of the cavity mode in the rotating frame, with $\delta_{cp}=\omega_{c}-\omega_{p}$. The operators $b_{m,j}$ and $b_{m,j}^\dagger$ are the bosonic annihilation and creation operators of the atoms where $m=0,1$ labels the legs of the ladder and $j$ the rungs of the ladder. The operator $n_{m,j}=b_{m,j}^\dagger b_{m,j}$ is the number operator. $L$ denotes the number of rungs of the ladder and the total number of bosons is $N$. The filling per site is defined as $\rho=N/(2L)$. $H_\|$ describes the tunneling of the atoms along the legs of the ladder with the tunneling amplitude $J_\|$. The term $H_{int}$ represents the repulsive on-site interaction of strength $U>0$. The term $H_{trap}$ represents an external harmonic trapping potential with $j_0=(L+1)/2$ which is typically present in nowadays experimental setups. The coupling between the atoms and the cavity field is described by $H_{ac}$, where a tunneling event along the rungs occurs by creation or annihilation of a cavity photon. The tunneling along the rungs with the spatially dependent phase imprint is represented by the operator $K_\perp$. In order to prevent a privileged direction of tunneling, the tunneling in each direction is coupled to both the creation and annihilation operators of the cavity field, using two pump laser beams \cite{DimerCarmichael2007}. The process has the amplitude $\hbar\tilde{\Omega}=\frac{\hbar\Omega_{p,1} g_0}{\omega_e-\omega_{p,1}} \phi_\| \phi_\perp$, where 
the effective parameters $\phi_\|$ and $\phi_\perp$ contain contributions of the overlap of the wavefunctions from neighboring sites and can be controlled by the geometry of the lattice \cite{CorinnaAmenehStefanPRL2016}. The two Raman processes are balanced due to the choice of the Rabi frequency for the second pump beam $\Omega_{p,2}= \Omega_{p,1}\frac{\omega_e-\omega_{p,2}}{\omega_e-\omega_{p,1}}$. 

Dissipative processes are present due to the imperfections of the cavity mirrors where losses of the cavity photons occur. The dissipative dynamics is approximated by a Lindblad master equation. The evolution of an operator $O$ is given by

\begin{align}
\label{eq:Lindblad}
& \pdv{t} O = \frac{i}{\hbar} \left[ H, O \right] + \mathcal{D}(O), 
\end{align}
with the dissipator $\mathcal{D}(O) =  \kappa \left( 2a^\dagger O a - O a^\dagger a - a^\dagger a O  \right)$, which gives the loss of cavity photons, via the imperfect mirrors.

\subsection{\label{sec:adiabiatic}Adiabatic elimination of the cavity field}
\label{2b}
In this section we will derive an effective model for the bosonic atoms, obtained from the adiabatic elimination of the cavity field \cite{RitschEsslinger2013, CorinnaAmenehStefanPRL2016, KollathBrennecke2016}. A full simulation of the problem would be interesting in order to investigate whether retardation effects of the cavity field might occur. However, here we approximate the cavity field with its steady state value, which fulfills the condition $\partial_t \langle a\rangle=0$. 
Using Eq. (\ref{eq:Lindblad}) the condition is given by

\begin{equation}\label{eq:dyn_photon}
i \partial_t \langle a\rangle=-\tilde{\Omega}\langle K_\perp+K_\perp^\dagger \rangle +(\delta_{cp}- i \kappa ) \langle a\rangle=0.
\end{equation}

This condition relates the expectation value of the directed rung tunneling to the value of the cavity field by
\begin{equation}\label{eq:alpha}
\alpha=\langle a \rangle = \frac{\tilde{\Omega}}{\delta_{cp}- i \kappa }\langle  K_\perp+K_\perp^\dagger\rangle.
\end{equation}

The model exhibits a $\mathbb{Z}_2$ symmetry, associated with the inversion of the sign of both the rung tunneling, $K_\perp+K_\perp^\dagger$, and the cavity field, $a+a^\dagger$. 
In an experiment this symmetry will be spontaneously broken such that each of the realizations of non-trivial steady state has the same magnitude of the expectation value $\abs{\langle a+a^\dagger\rangle}$, however, with a spontaneously chosen sign. 
We choose without loss of generality $\langle K_\perp+K_\perp^\dagger\rangle>0$ and even find for the considered phases that we can choose  $\langle K_\perp\rangle>0$.

The equations of motion of the bosonic operators read

\begin{eqnarray}
\label{eq:dyn_boson}
i \hbar \partial_t \langle b_{0,j} \rangle=-J_\|\langle b_{0,j+1} + b_{0,j-1} \rangle-U\langle b_{0,j}(1-n_{0,j}) \rangle\nonumber \\-\hbar\tilde{\Omega}\langle a+a^\dagger\rangle e^{i\varphi j} \langle b_{1,j} \rangle +\frac{4V_{trap}}{L^2}(j-j_0)^2 \langle b_{0,j} \rangle~~~ \nonumber\\
i \hbar \partial_t \langle b_{1,j} \rangle= -J_\|\langle b_{1,j+1} + b_{1,j-1} \rangle-U\langle b_{1,j}(1-n_{1,j}) \rangle\nonumber \\-\hbar\tilde{\Omega}\langle a+a^\dagger\rangle e^{-i\varphi j} \langle b_{0,j} \rangle +\frac{4V_{trap}}{L^2}(j-j_0)^2 \langle b_{1,j} \rangle.~~
\end{eqnarray}
In the derivation of the equations we have used a mean-field decoupling of the atomic and cavity degrees of freedom.

We substitute the expectation value for the cavity field, Eq.~(\ref{eq:alpha}), into the equations of motion of the bosonic operators, Eq.~(\ref{eq:dyn_boson}). The dynamics given by the obtained set of equations can be described by an effective Hamiltonian for the atoms 

\begin{align}
&H=H_\parallel+H_\perp+H_{int}+H_{trap} \label{eq:eff_ham}
\\
&H_\parallel=-J_\|\sum_{j,m=0,1} (b_{m,j}^\dagger b_{m,j+1} + b_{m,j+1}^\dagger b_{m,j})\nonumber\\
&H_\perp= -J_\perp (K_\perp + K_\perp ^\dagger) \nonumber\\
&H_{int}=\frac{U}{2} \sum_{j,m=0,1} n_{m,j}(n_{m,j}-1)\nonumber\\
&H_{trap}=\frac{4 V_{trap}}{L^2}\sum_{j,m=0,1}\qty(j-j_0)^2n_{m,j}\nonumber
\end{align}
and typically the ground state of this effective model corresponds to the steady state of the equations of motion.
The parameter $J_\perp$, which represents the rung tunneling amplitude, has to be determined self-consistently as it depends on the expectation value of $\langle K_\perp \rangle$, $J_\perp=A\langle K_\perp \rangle$, with $A= \frac{4\hbar\tilde{\Omega}^2\delta_{cp}}{\delta_{cp}^2 +\kappa^2}$. The self-consistency condition reflects the global nature of the coupling of the atoms to the cavity by the presence of the global rung tunneling.
In order to get a non-trivial solution ($\alpha\neq 0$) of the self-consistency condition we must require $A>0$, which implies $\delta_{cp}>0$.

\subsection{\label{sec:stability}Stability analysis}
\label{2c}

The non-trivial self-consistent solution(s) derived from the effective Hamiltonian, Eq.~(\ref{eq:eff_ham}), might not be stable. Thus, in this section we derive a stability condition for our model using pertubations around the steady state \cite{RitschEsslinger2013}. We follow the notations introduced in Ref.~\cite{Tian2016}.

Starting from the equation of motion for the cavity field, given by Eq.~(\ref{eq:dyn_photon}), we introduce the averages of the coordinate and momentum quadratures of the cavity field, $x_a=\langle a+a^\dagger\rangle$ and $p_a=-i\langle a-a^\dagger\rangle$. Using Eq.~(\ref{eq:dyn_photon}) and its conjugate we have

\begin{align}
\label{eq:quadratures}
&\frac{\partial}{\partial t} x_a=-\kappa x_a+\delta_{cp}p_a\\
&\frac{\partial}{\partial t} p_a=-\delta_{cp}x_a-\kappa p_a+4\tilde{\Omega} \langle K_\perp \rangle. \nonumber
\end{align}

The stationary solutions of these equations, which satisfy the relations $\partial_tx_a^{(s)}=0$ and  $\partial_tp_a^{(s)}=0$ are

\begin{align}
\label{eq:statsol}
&x_a^{(s)}=\frac{4\delta_{cp}\tilde{\Omega}\langle K_\perp \rangle^{(s)}}{\delta_{cp}^2+\kappa^2},\\
&p_a^{(s)}=\frac{4\kappa\tilde{\Omega}\langle K_\perp \rangle^{(s)}}{\delta_{cp}^2+\kappa^2}, \nonumber
\end{align}
where the average of the directed tunneling $\langle K_\perp \rangle$ computed in the ground state of the effective model, Eq.~(\ref{eq:eff_ham}), can have a nonlinear dependence on the stationary coordinate quadrature $x_a^{(s)}$.

We will consider linear fluctuations around the stationary solutions, i.e., $x_a=x_a^{(s)}+\tilde{x}_a$ and $p_a=p_a^{(s)}+\tilde{p}_a$, and also linearize the average of the directed tunneling in terms of the fluctuations

\begin{equation}
\label{eq:kperlin}
\langle K_\perp \rangle=\langle K_\perp \rangle^{(s)}+\frac{d\langle K_\perp \rangle^{(s)}}{dx_a^{(s)}}\tilde{x}_a\;,
\end{equation}
where $\langle K_\perp \rangle^{(s)}$ is the value of the directed rung tunneling corresponding to the stationary solution $x_a^{(s)}$.
From Eqs.~(\ref{eq:quadratures}) and (\ref{eq:kperlin}) we can derive a set of differential equations for the fluctuations

\begin{align}
\label{eq:fluctuations}
&\frac{\partial}{\partial t} \tilde{x}_a=-\kappa \tilde{x}_a+\delta_{cp}\tilde{p}_a\\
&\frac{\partial}{\partial t} \tilde{p}_a=\left(-\delta_{cp}+4\tilde{\Omega}\frac{d\langle K_\perp \rangle^{(s)}}{dx_a^{(s)}}\right)\tilde{x}_a-\kappa \tilde{p}_a. \nonumber
\end{align}
The eigenvalues of the Jacobian of this set of differential equations are given by

\begin{equation}
\label{eq:eigenvalue}
\lambda_{\pm}=-\kappa\pm\sqrt{\left(-\delta_{cp}^2+4\delta_{cp}\tilde{\Omega}\frac{d\langle K_\perp \rangle^{(s)}}{dx_a^{(s)}}\right)}.
\end{equation}
The stable stationary solutions are the ones for which the eigenvalues have a negative real part. 
Thus the stability condition for the system with $\delta_{cp}>0 $ is
\begin{equation}
\label{eq:condition1}
\frac{\delta_{cp}^2+\kappa^2}{4\delta_{cp}\tilde{\Omega}}>\frac{d\langle K_\perp \rangle^{(s)}}{dx_a^{(s)}}.
\end{equation}
This condition can be rewritten in a form that we can easily use in our model, using the relation $J_\perp^{(s)}=\hbar \tilde{\Omega}x_a^{(s)}$
\begin{equation}
\label{eq:condition2}
\frac{d\langle K_\perp \rangle^{(s)}}{dJ_\perp^{(s)}}<\frac{1}{A}.
\end{equation}
Let us comment that this stability condition means that if the pump strength A is slightly increased, the slope of the line $J_\perp/ A$ will decrease. 
If we start from a stable solution, the corresponding slightly changed solution lies at larger values of  $\langle K_\perp  \rangle$ and $J_\perp$, whereas if we start from an unstable solution its value decreases, moving closer to the trivial solution of $\langle K_\perp \rangle=0$ and $J_\perp=0$. 

\subsection{\label{sec:selfconst}Self-consistent solution}

We determine the stable steady states of the coupled cavity-atomic model using three steps: First we perform a ground-state search using matrix product state methods (DMRG) for the effective model Eq.~(\ref{eq:eff_ham}),
for fixed flux $\varphi$, filling $\rho$ and on-site interaction $U/J_\parallel$, while varying the rung tunneling amplitude $J_\perp/J_\parallel$. We compute the expectation value $\langle K_\perp\rangle$ obtaining its dependence on $J_\perp/J_\parallel$. In the second step we find the solution of the self-consistent problem reformulated as
\begin{equation}
\label{eq:condition3}
\langle K_\perp \rangle=\frac{J_\|}{A}~\frac{J_\perp}{J_\|}.
\end{equation}
The left-hand side contains the nonlinear behavior of $\langle K_\perp \rangle$ given by the effective model and the right-hand side is a linear function with slope $J_\|/A$. Graphically, the solutions are the crossings of the two curves. The last step is to work out which of these solutions are stable. The stability condition, Eq.~(\ref{eq:condition2}), tells us that we need to compare the slope of the two curves at their intersection point(s). The solution is stable if the slope of $\langle K_\perp \rangle$ is smaller than the slope of the linear function. The results will be shown in Sec.~\ref{sec:results}.

\subsection{\label{sec:effectiveH}Properties of the effective Hamiltonian}

\begin{figure}[hbtp]
\centering
\includegraphics[width=.5\textwidth]{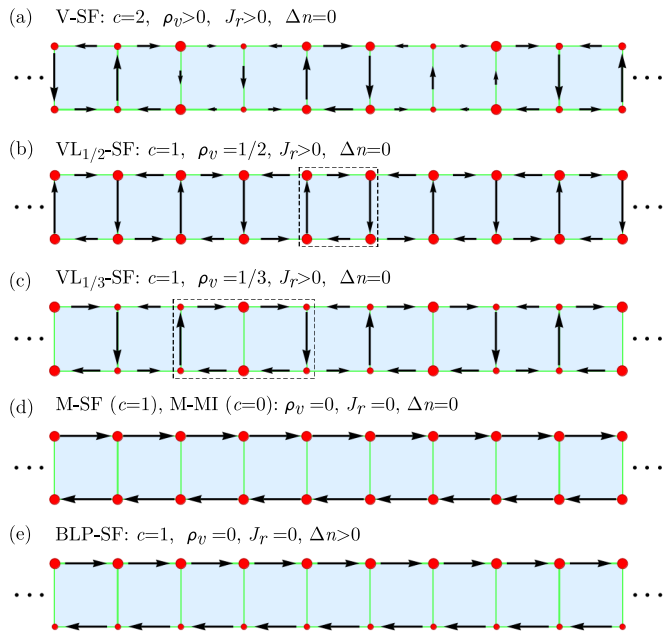}
\caption{Current patterns and on-site density for some of the different phases of the effective model, (a) vortex liquid (V-SF) phase with a vortex density $1/3<\rho_v<2/5$, (b)-(c) vortex lattices (VL$_{1/2,1/3}$-SF) with (b) $\rho_v=1/2$, and (c) $\rho_v=1/3$, (d) the Meissner phase (M-SF, M-MI), and (e) the biased-ladder phase (BLP-SF). The length of the arrows is proportional to the strength of the local currents and the size of the red circles scales with the on-site density (DMRG data). The dashed rectangles represent the unit cell for the vortex lattice phases. 
 }
\label{fig:current_patterns}
\end{figure}

The effective model, Eq.~(\ref{eq:eff_ham}), in the abscence of a trapping potential, has been studied as a stand-alone model describing bosonic two-leg ladders with repulsive contact interaction in the presence of a uniform, external gauge field.  DMRG and bosonization studies have explored the phase diagram \cite{OrignacGiamarchi2001,TokunoGeorges2014,PiraudSchollwock2015, GreschnerVekua2015,GreschnerVekua2016, Dhar2012, Dhar2013, WeiMuller2014, UchinoTokuno2015, Uchino2016, PetrescuHur2015, StrinatiMazza2016, PetrescuHur2017, DioChiofalo2015, OrignacChiofalo2016, OrignacPalo2017}, observing Meissner, vortex and vortex lattices phases on top of superfluid or Mott-insulating states. The phase transitions that occur in this model take place in two sectors. These sectors correspond, in the limit of weak coupling (i.e. $J_\perp/J_\|\ll 1$), to the symmetric ($b_{0,j}+b_{1,j}$) and antisymmetric ($b_{0,j}-b_{1,j}$) combination of the operators on the two legs. The Mott insulator to superfluid transition is located in the symmetric sector, while the Meissner to vortex phase transitions occur in the antisymmetric sector.

One quantity which gives information on the phases is the central charge $c$, which can be interpreted as the number of gapless modes. The central charge can be extracted from the scaling of the von Neumann entanglement entropy $S_{vN}(l)$ of an embedded subsystem of length $l$ in a chain of length $L$. For open boundary conditions the entanglement entropy for the ground state of gapless phases scales as \cite{VidalKitaev2003,CalabreseCardy2004, HolzeyWilczek1994}

\begin{equation}
\label{eq:entropy}
S_{vN}=\frac{c}{6}\log\left(\frac{L}{\pi}\sin\frac{\pi l}{L}\right)+s_1,
\end{equation}
where $s_1$ is a non-universal constant and we have neglected  
logarithmic corrections \cite{AffleckLudwig1991} and oscillatory terms \cite{LaflorencieAffleck2006} due to the finite size of the system. 

One can distinguish between the superfluid and Mott-insulating phases by the decay of the single particle correlation, here we use $\langle b^\dagger_{m,L/4} b_{m,L/4+d}+\text{H.c.} \rangle$, with distance $d$. The correlations decay algebraically with distance in a superfluid state and exponentially in a Mott-insulating state \cite{Giamarchibook}. 

Important characteristics of the quantum chiral phases (Meissner, vortex liquid, or vortex lattice) can be inferred from the configurations of the local currents, which is experimentally accessible in the bosonic ladder system \cite{AtalaBloch2014, StuhlSpielman2015}. We define the local currents on the leg $j^\|_{m,j}$ and the rung $j^\perp_{j}$, respectively, as

\begin{align}
\label{eq:localcur}
&j^\|_{m,j} = -i J_\|(b_{m,j}^\dagger b_{m,j+1} -\text{H.c.}),\nonumber \\
&j^\perp_{j} = -i J_\perp (e^{i\varphi j}b_{1,j}^\dagger b_{0,j} -\text{H.c.}). 
\end{align}
In addition to the local currents, the chiral current $J_c$ and the average rung current $J_r$ are of interest and defined as

\begin{align}
\label{eq:cur}
&J_c = \frac{1}{2L} \sum_j \langle j^\|_{0,j} - j^\|_{1,j} \rangle,\nonumber\\ 
&J_r=\frac{1}{L} \sum_j |\langle j^\perp_{j}\rangle|.
\end{align}
 All non-trivial phases considered here have a finite chiral current. The Meissner phases are characterized by a vanishing average rung current, $J_r=0$, and the vortex and vortex lattice phases by non-zero rung currents, $J_r>0$. An additional quantity, the vortex density is defined as the inverse of the vortex length, $\rho_v=l_v^{-1}$. Where $l_v$ is the typical size of the vortices and we extract it from the Fourier transform of the local rung current configurations $\langle j_r^\perp \rangle$. From the value of $\rho_v$ one identifies the periodicity of the vortex lattice phases. In computing the observables we only use the central region, $j\in[L/4,3L/4]$, of the ladder, in order to reduce the influence of boundary effects. 

In the following we will describe the phases that appear as steady states for the sets of parameters that will be considered in Sec.~\ref{sec:results}. A summary is given in Fig.~\ref{fig:current_patterns}. The vortex superfluid phase (V-SF), with the local current and density pattern depicted in Fig.~\ref{fig:current_patterns}(a), has two gapless modes, thus it has a central charge $c=2$. It is also characterized by a finite vortex density, $\rho_v$, which is incommensurate with the ladder.

The existence of vortex lattice phases has been demonstrated in bosonic ladders by Refs.~\cite{Dhar2012, Dhar2013, GreschnerVekua2015, GreschnerVekua2016}, for different vortex densities. 
In Fig.~\ref{fig:current_patterns}(b)-(c) we have represented the current and density pattern for the vortex lattice superfluid phases ($\text{VL}_{\rho_v}\text{-SF}$) with $\rho_v=1/2$ and $1/3$. Where the vortices are commensurate with the ladder. Both of these two phases have a central charge, $c=1$, since the symmetric sector is gapless and the antisymmetric one is gapped. Due to the spontaneous symmetry breaking of the translation symmetry in the vortex lattices, the unit cell becomes $q$-fold enlarged, with $q=2$ and $3$ in these cases. This can lead to a change in the sign of the chiral current \cite{GreschnerVekua2015}. 

In Fig.~\ref{fig:current_patterns}(d) we depicted a Meissner phase, which can be a Meissner superfluid (M-SF) or a Meissner Mott insulator (M-MI). Both of these phases have vanishing currents on the rungs in the bulk of the system and a finite chiral current. The distinction between the two phases can be made by calculating the central charge. The Meissner superfluid has a gapless symmetric sector, while the Meissner Mott insulator is totally gapped.

The biased-ladder phase (BLP) \cite{WeiMuller2014,UchinoTokuno2015, Uchino2016,GreschnerVekua2015,GreschnerVekua2016}, Fig.~\ref{fig:current_patterns}(e), breaks the discrete $\mathbb{Z}_2$ symmetry associated with the inversion of the two legs of the ladder and the sign of the flux. The characteristic signature of this phase is that the density is higher on one of the legs, compared to the other. We will use the density imbalance, $\Delta n$, to identify this phase. The imbalance is defined as

\begin{equation}
\label{eq:imbalance}
\Delta n=\frac{1}{2L} \abs\Big{\sum_j \langle n_{0,j}-n_{1,j} \rangle}.
\end{equation}
The biased ladder phase has a gapless mode in the symmetric sector, and a vanishing rung current.

As the biased ladder phase state spontaneously breaks the $\mathbb{Z}_2$ symmetry between the two legs, it results in a twofold degenerate ground state. The DMRG ground state search can choose any state within this degenerate subspace, resulting in an arbitrary value of the density imbalance between zero and the maximal value. In the following we will present the procedure employed in order to obtain the ground state with maximal value of $\Delta n$. After numerically obtaining one of the ground states, $\ket{\psi_1}$, we compute the orthogonal wavefunction, $\ket{\psi_2}$, which is degenerate in energy with $\ket{\psi_1}$.  For this, the second ground state search is performed enforcing the orthogonality by the following Hamiltonian

\begin{equation}
\label{eq:Horth}
\tilde{H}=H+w\ket{\psi_1}\bra{\psi_1},
\end{equation}
where $H$ is the Hamiltonian given in Eq.~(\ref{eq:eff_ham}) and the weight $w>0$ introduces an energy penalty for any finite overlap with $\ket{\psi_1}$ and guarantees the orthogonality. We use a typical value of $w=10$ and checked consistency with higher values up to $w=100$.
As an initial state for the search we use a state with an inverse density imbalance compared to the state $\ket{\psi_1}$. Having obtained an orthonormal basis in the ground state manifold, we move on to construct the superposition of these two states, $\ket{\psi_\lambda}=\lambda\ket{\psi_1}+\sqrt{1-\lambda^2}\ket{\psi_2}$ with $\lambda\in[0,1]$, that has the maximum density imbalance $\Delta n$. Compared to the induced symmetry breaking by an externally applied potential, we found this method to more reliably identify the imbalanced phase. 

The described states do not represent an exhaustive list of the possible phases of the ground-state of the bosonic ladder. We focused on the ones we can stabilize dynamically in the cavity, for the considered parameters. Beside the mentioned phases there exists numerical evidence for vortex and vortex lattice Mott insulators \cite{PiraudSchollwock2015, GreschnerVekua2015}, vortex lattice superfluid with $\rho_v=1/4$ and charge density waves \cite{GreschnerVekua2016}, also Laughlin states have been proposed \cite{PetrescuHur2015,StrinatiMazza2016, PetrescuHur2017}. 

\subsection{\label{sec:method}DMRG Method}

The results presented in this work were obtained using a finite-size density matrix renormalization group (DMRG) algorithm in the matrix product state form \cite{White1992, Schollwock2005, Schollwock2011, Hallberg2006, Jeckelmann2008}, using the ITensor Library \cite{itensor}. We simulate the presented model Eq.~(\ref{eq:eff_ham}) typically for a ladder with $L=120$ rungs and with the bond dimension up to 1500 in the matrix product state representation. We checked the converge of the method for different system sizes and bond dimensions. Since we are dealing with finite interactions the local Hilbert space of bosons is infinite, thus a cutoff for its dimension is needed. We use a maximal dimension of five bosons per site, where the higher cutoff of six bosons per site gives consistent results. 

\section{\label{sec:results}Results}

\begin{figure}[hbtp]
\centering
\includegraphics[width=.5\textwidth]{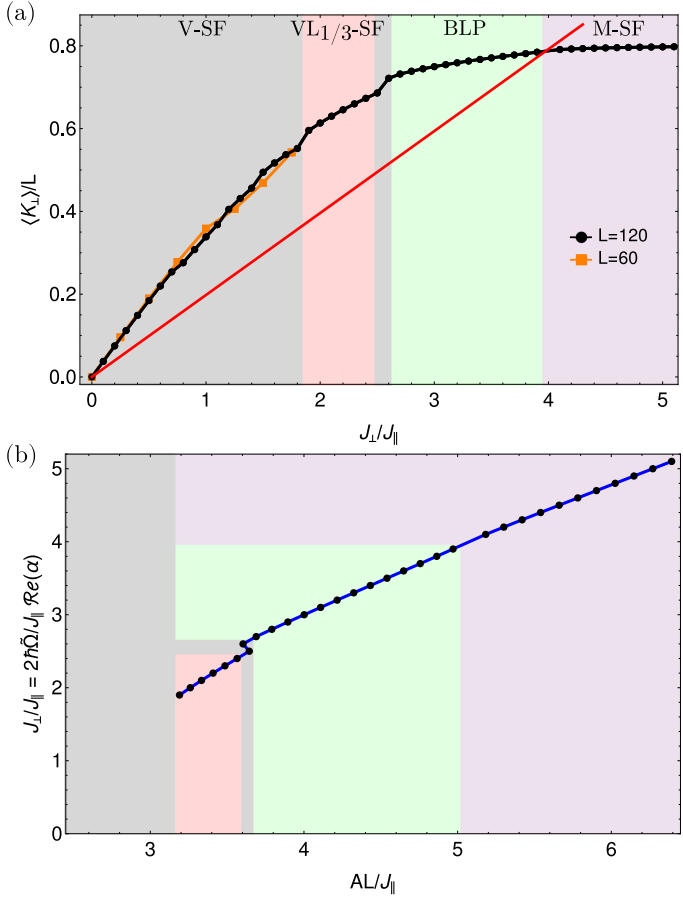}
\caption{(a) Graphical interpretation of the self-consistency condition for the parameters $\varphi=0.8 \pi$, $\rho=0.8$, $U=1J_\|$, and $V_{trap}=0$. The directed rung tunneling  $\langle K_\perp \rangle/L$ is represented for two system sizes, $L=120$ and $L=60$. The straight (red) line represents the right-hand side of the self-consistency condition, which is a linear function with slope $\frac{J_\|}{AL}$. The crossings of the two curves give the solutions of the self-consistency condition.  (b) The solutions $J_\perp/J_\|$ of the self-consistency equation  which are proportional to the cavity field $ Re(\alpha)$ versus the pump strength $AL/J_\|$. The filled colored areas represent the extent of different phases. In the grey area the stability of the solutions is not clear for all system sizes.
 }
\label{fig:kpersol08}
\end{figure}

\subsection{\label{sec:parameters1}Steady state diagram at flux $\varphi=0.8 \pi$, \\filling $\rho=0.8$ and on-site interaction $U=1J_\|$ in a homogeneous system}
\label{3b}

\begin{figure}[!hbtp]
\centering
\includegraphics[width=.5\textwidth]{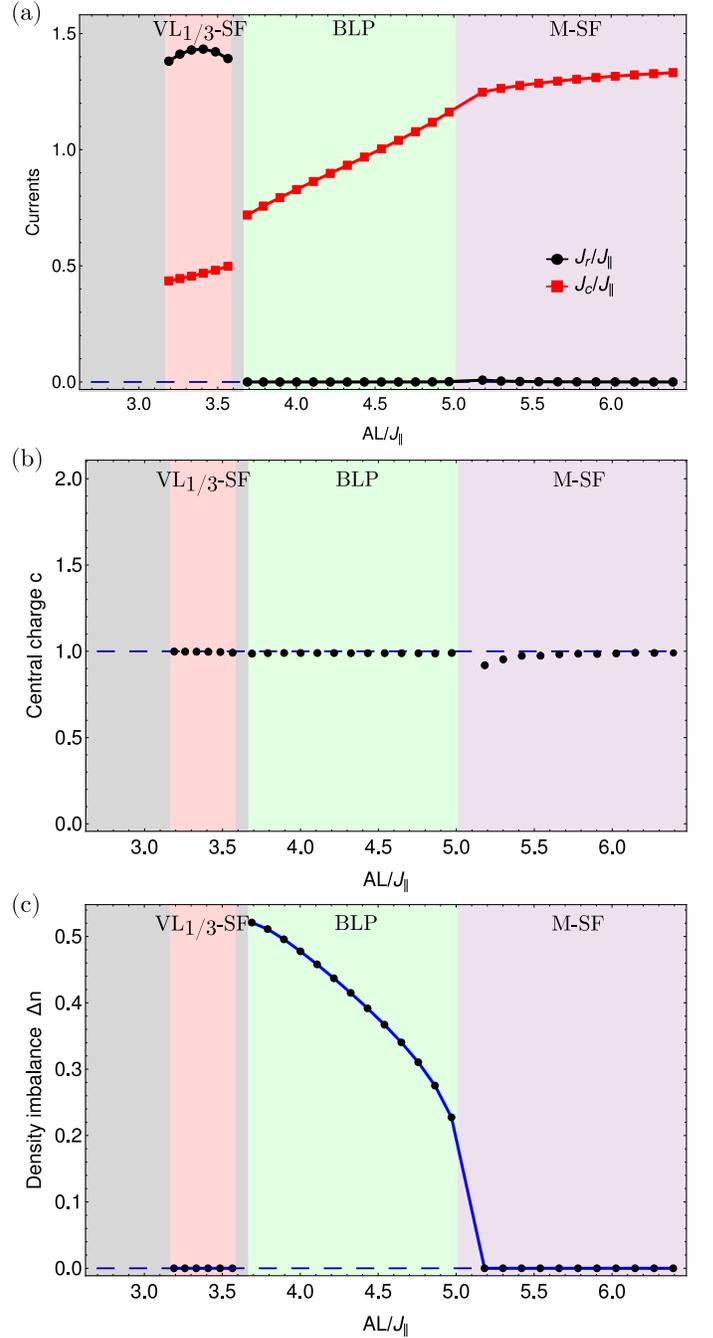}
\caption{(a) Chiral current $J_c$ and average rung current $J_r$, (b) central charge $c$, and (c) the density imbalance between the two legs of the ladder as a function of the pump strength $AL/J_\|$ in the stable regions for the parameters $\varphi=0.8\pi$, $\rho=0.8$, $U=1J_\|$, and $V_{trap}=0$. The central charge is extracted fitting the scaling of entanglement entropy. The error bars of the fit are smaller than the symbols used. Dashed horizontal lines indicate the constant value 0 or 1, as a guide to the eye. 
 }
\label{fig:observables08}
\end{figure}

In this subsection we solve the self-consistency condition and identify the steady states which can be stabilized as we vary the pump strength $A$, for flux $\varphi=0.8\pi$, filling $\rho=0.8$ and on-site interaction $U=1J_\|$ in a homogeneous system. For this case the dynamic stabilization of $\text{VL}_{1/3}\text{-SF}$, biased ladder superfluid, and Meissner superfluid states is possible. As mentioned, the first step in tackling the self-consistency condition, Eq.~(\ref{eq:condition3}), is to calculate the expectation value of the directed rung tunneling $\langle K_\perp \rangle/L$. The intersections of this curve with the linear function $\frac{J_\|}{AL}$ give the solutions of the self-consistency condition, shown in Fig.~\ref{fig:kpersol08}. From Fig.~\ref{fig:kpersol08}(a) we can also infer the stability of the solutions, using Eq.~(\ref{eq:condition2}), by comparing the slopes of the two curves. If the derivative of  $\langle K_\perp \rangle/L$ is less than the slope of the linear function, the solution is stable. We evaluate the derivative numerically by computing the left and right finite derivatives with the help of the two adjacent points. We consider a solution stable if both the left and right finite derivatives satisfy the condition given by Eq.~(\ref{eq:condition2}).
The expectation value of the directed rung tunneling $\langle K_\perp \rangle/L$ has a concave curvature, with some additional substructure especially close to the phase transitions. For small values of $J_\perp$, $\langle K_\perp \rangle/L$ has a strong dependence on the size of the system, as we observe from Fig.~\ref{fig:kpersol08}(a) and thus the stability in this region depends crucially on the size of the system. In the following we will concentrate on the states which are stable for all considered system sizes. The non-trivial stable solutions are plotted in Fig.~\ref{fig:kpersol08}(b), where we have always chosen the solution $J_\perp>0$. Not shown are the corresponding solutions at the inverse of the value which exist due to the $\mathbb{Z}_2$ symmetry of the model. 

\begin{figure}[hbtp]
\centering
\includegraphics[width=.5\textwidth]{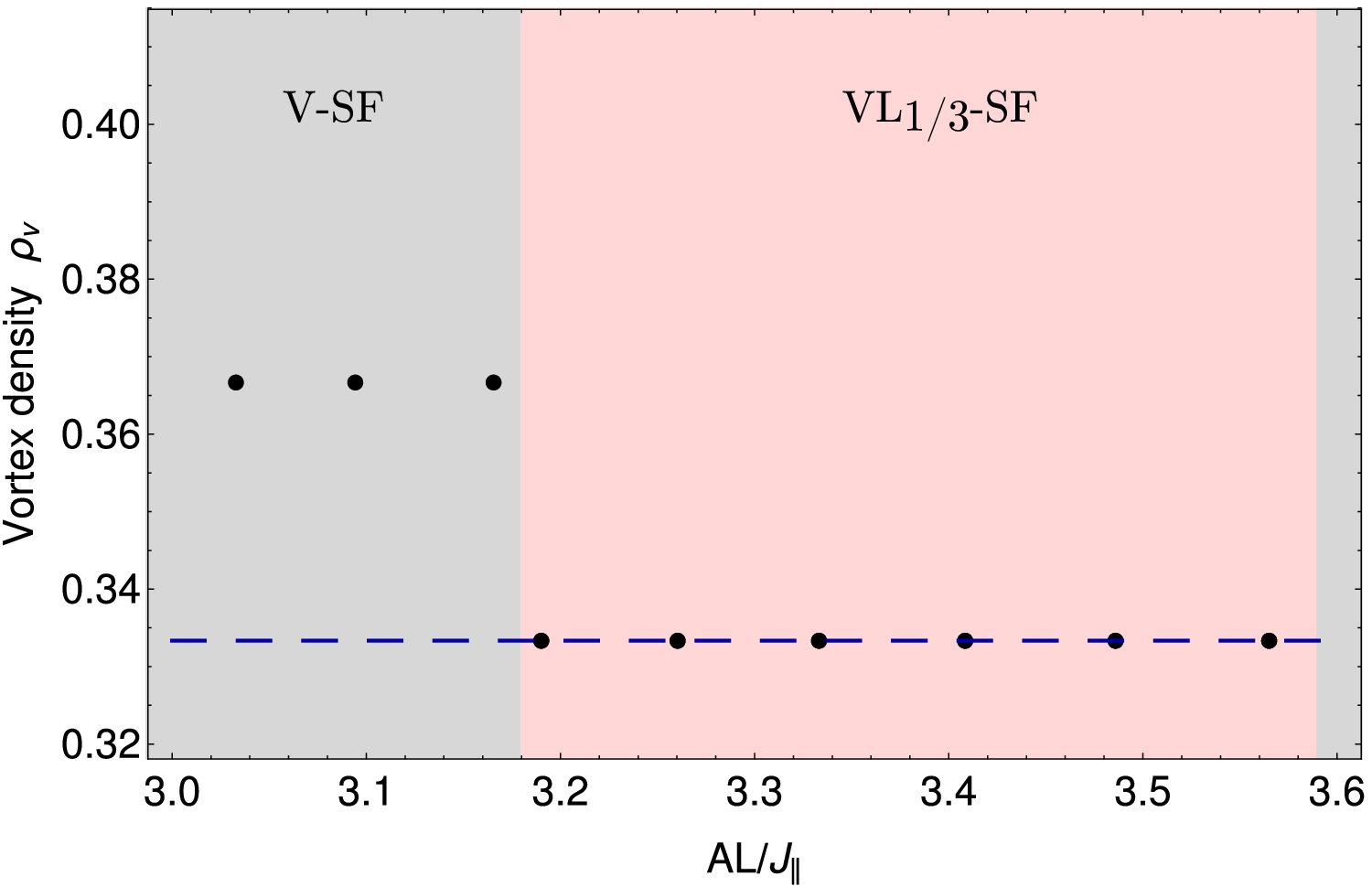}
\caption{The vortex density $\rho_v$ as a function of the pump strength $AL/J_\|$, where the rung current is finite, for the parameters $\varphi=0.8\pi$, $\rho=0.8$, $U=1J_\|$, and $V_{trap}=0$. The vortex density has the value $\rho_v=1/3$ for $3.19J_\|/L\lesssim A\lesssim 3.59J_\|/L$, which represents the $\text{VL}_{1/3}\text{-SF}$ phase. Dashed horizontal line indicates the constant value $1/3$. 
 }
\label{fig:vortexd08}
\end{figure}

One can observe from Fig.~\ref{fig:kpersol08}(b) that we have (system size independent) non-trivial stable solutions  with a finite occupation of the cavity field  above the pump strength $A\approx3.19J_\|/L$.

\begin{figure}[hbtp]
\centering
\includegraphics[width=.5\textwidth]{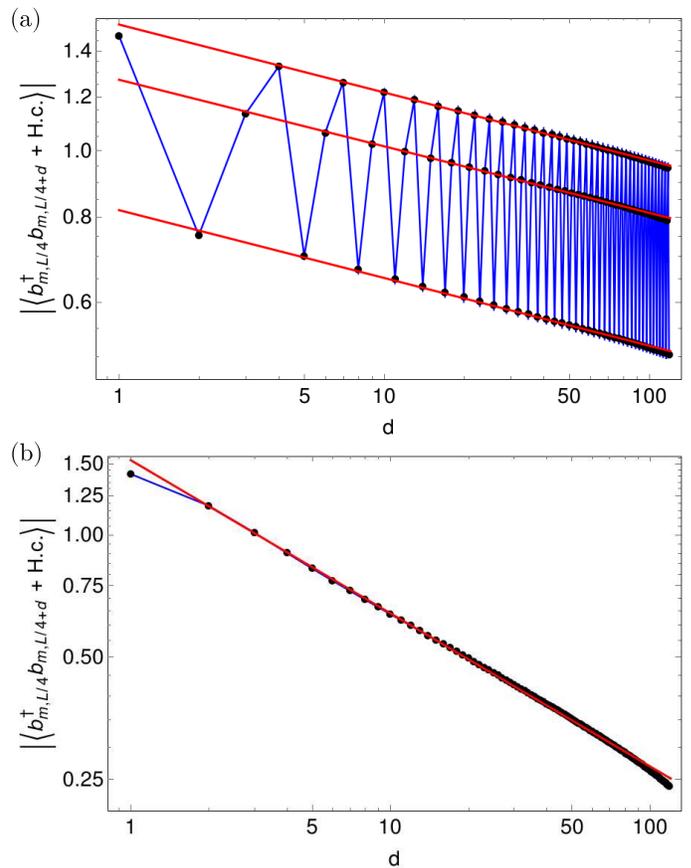}
\caption{The absolute value of the single particle correlations, $|\langle b^\dagger_{m,L/4} b_{m,L/4+d}+\text{H.c.}\rangle|$, along one of the legs represented in a logarithmic plot for the parameters $L=240$, $\varphi=0.8\pi$, $\rho=0.8$, $U=1J_\|$, and $V_{trap}=0$, in the (a) $\text{VL}_{1/3}$ state, for $A=3.44 J_\|/L$, (b) Meissner state, for $A=6.26 J_\|/L$. The correlations show an algebraic decay with distance, which signals the superfluid phase. The straight (red) lines are fits of the function $\propto x^{-\alpha}$, where the fit parameter is (a) $\alpha=0.098\pm 0.001$, (b) $\alpha=0.377\pm 0.001$. 
 }
\label{fig:corr08}
\end{figure}

The next step consists in identifying the properties of the steady states that correspond to the stable solutions. We have to analyze the behavior of the observables characterizing these states (Fig.~\ref{fig:observables08}), as explained in Seq.~\ref{sec:effectiveH}. 
We find a vortex state, with finite rung currents, for $3.19J_\|/L\lesssim A\lesssim 3.59J_\|/L$. We extract the central charge by fitting Eq.~(\ref{eq:entropy}) to the numerically computed  von Neumann entropy. The central charge has a value of $c\approx 1$ in this region, which points, together with the finite rung current, towards a lattice vortex superfluid state. The superfluid nature can also be confirmed by an algebraic decay of the single particle correlation function along the ladder (see Fig.~\ref{fig:corr08}(a)). The algebraic decay is modulated by a periodic function, with the period of three lattice sites. From the vortex density (Fig.~\ref{fig:vortexd08}), we identify the $\text{VL}_{1/3}\text{-SF}$ state corresponding to this modulation. 

In the region $3.66J_\|/L\lesssim A\lesssim 5.01J_\|/L$, the density imbalance between the two legs acquires finite values with a sharp onset at the lower boundary. This signals the biased ladder state. 
 Our numerical data is not precise enough in order to identify the transition between $\text{VL}_{1/3}\text{-SF}$ and the biased ladder states in the region $3.59J_\|/L\lesssim A\lesssim 3.66J_\|/L$. Whereas our numerical data (Fig.~\ref{fig:kpersol08}(b)) suggests multiple solutions for the same value of the pump strength, from our numerical resolution we cannot decide on their stability condition. In the biased ladder state, the currents along the rungs of the ladder are suppressed, which is consistent with the numerical data from Fig.~\ref{fig:observables08}(a). Additionally, the charge remains constant around $c\approx 1$. 

For large values of the pump strength $A\gtrsim 5.01J_\|/L$, the density imbalance vanishes (see Fig.~\ref{fig:observables08}(c)). The chiral current saturates, while the currents on the rungs are still suppressed (see Fig.~\ref{fig:observables08}(a)). The single particle correlation function decays algebraically. Considering these previous findings together with the fact that the state has one gapless mode ($c\approx 1$), the state above $A\gtrsim 5.01J_\|/L$ is a Meissner superfluid. 

Thus, for the parameters $\varphi=0.8\pi$, $\rho=0.8$ and $U=1J_\|$ the dynamical stabilization of a vortex lattice superfluid with $\rho_v=1/3$, a biased ladder superfluid and a Meissner superfluid states is possible.

\begin{figure}[hbtp]
\centering
\includegraphics[width=.5\textwidth]{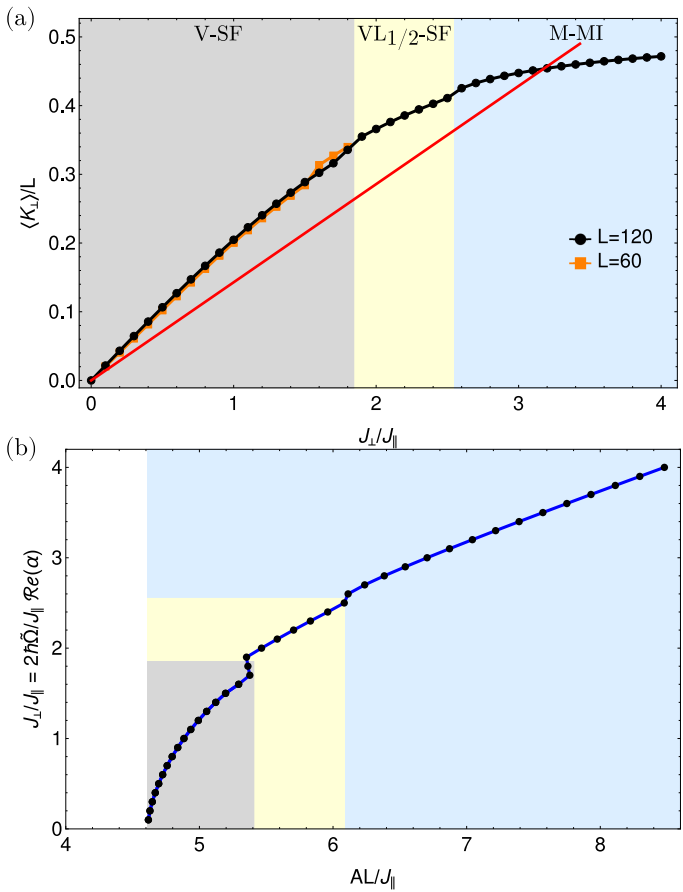}
\caption{(a) The expectation value of the directed rung tunneling  $\langle K_\perp \rangle/L$ for the parameters $\varphi=0.9\pi$, $\rho=0.5$, $U=1.5J_\|$, and $V_{trap}=0$, for two system sizes, $L=120$ and $L=60$. The straight (red) line represents the right-hand side of the self-consistency condition, which is a linear function with the slope $\frac{J_\|}{AL}$. The crossings of the two curves give the solutions of the self-consistency equation.  (b) The solutions $J_\perp/J_\|$ of the self-consistency equation  which are proportional to the cavity field $2\hbar\tilde{\Omega} Re(\alpha)$ versus the pump strength $AL/J_\|$ for $L=120$. The filled colored areas represent the extent of different phases. In the grey area the stability of the solutions is not clear for all system sizes.
 }
\label{fig:kpersol09}
\end{figure}

\subsection{\label{sec:parameters2}Steady state diagram at flux $\varphi=0.9\pi$, \\filling $\rho=0.5$ and on-site interaction $U=1.5J_\|$ in a homogeneous system}
\label{3c}

\begin{figure}[hbtp]
\centering
\includegraphics[width=.5\textwidth]{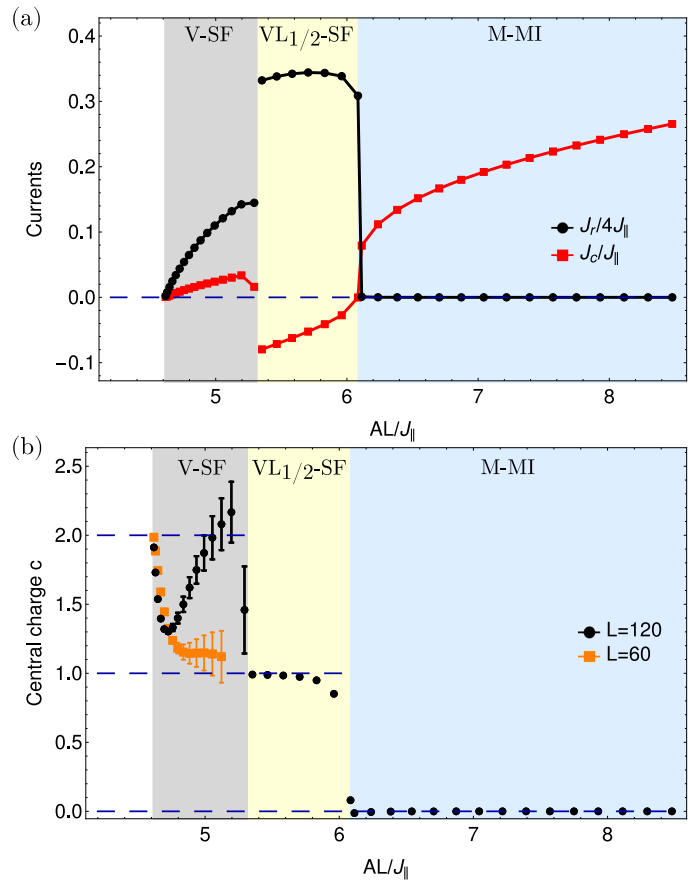}
\caption{(a) Chiral current $J_c$ and average rung current $J_r$ as a function of the pump strength $AL/J_\|$, (b) central charge $c$, computed from the scaling of entanglement entropy, for the parameters $\varphi=0.9\pi$, $\rho=0.5$, $U=1.5J_\|$, and $V_{trap}=0$. The errorbars represented the fit error. In the vortex superfluid region we represented the value of the central charge for two system sizes, $L=120$ and $L=60$, which shows a strong size dependent behavior. Dashed horizontal lines indicate the constant value 0, 1 or 2, as a guide to the eye.}
\label{fig:observables09}
\end{figure}

\begin{figure}[thbp]
\centering
\includegraphics[width=.5\textwidth]{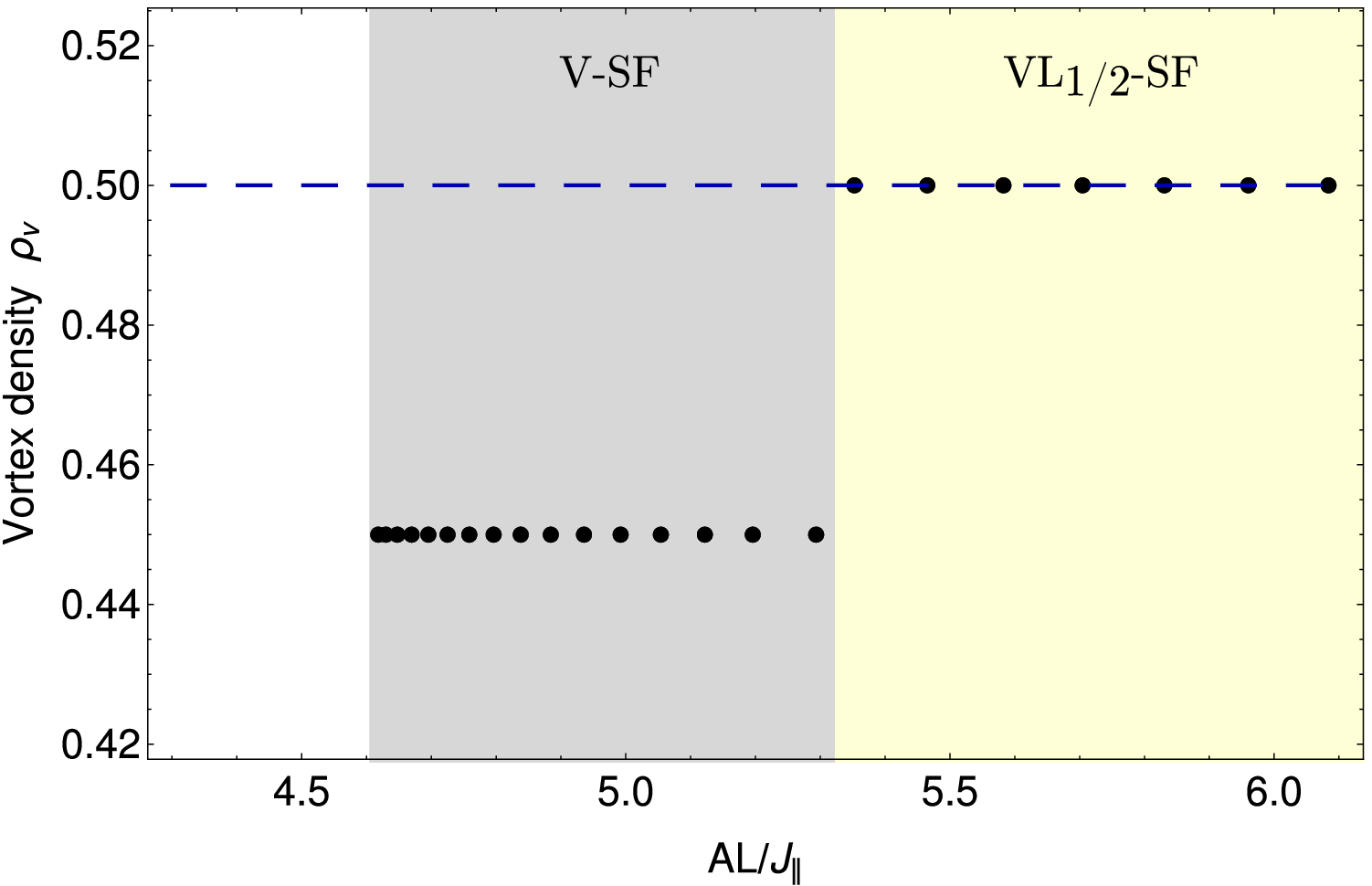}
\caption{The vortex density $\rho_v$ as a function of the pump strength $AL/J_\|$, where the rung current is finite for the parameters $\varphi=0.9\pi$, $\rho=0.5$, $U=1.5J_\|$, and $V_{trap}=0$. The vortex density has the value $\rho_v=1/2$ for $5.3J_\|/L\lesssim A\lesssim 6.1J_\|/L$, which represents the $\text{VL}_{1/2}\text{-SF}$ phase. Dashed horizontal line indicates the constant value $1/2$. 
 }
\label{fig:vortexd09}
\end{figure}

In the following, we will present which steady states can be dynamically organized for the parameters ${\varphi=0.9\pi}$, $\rho=0.5$ and $U=1.5J_\|$ in a homogeneous system. We show that a dynamical stabilization of a vortex superfluid, a $\text{VL}_{1/2}\text{-SF}$, and a Meissner Mott-insulating state is possible.  In Fig.~\ref{fig:kpersol09}(a) the expectation value of the directed rung tunneling $\langle K_\perp \rangle/L$ has been plotted for $L=120$ and $L=60$. For small $J_\perp$, corresponding to the vortex phases, $\langle K_\perp \rangle$ increases rapidly with $J_\perp$, whereas for large $J_\perp$, in the Meissner phase, $\langle K_\perp \rangle$ has a slow increase. In the regime where the two system sizes agree (for $J_\perp>1.8$), $\langle K_\perp \rangle$ is monotonic and concave, with small jumps near the phase transitions. 
 From the intersection of this curve with a line with the slope $\frac{J_\|}{AL}$ and checking the stability condition (\ref{eq:condition2}), we compute the stable solutions plotted in Fig.~\ref{fig:kpersol09}(b). 

For the size $L=120$ a stable solution can be found in the entire parameter regime shown, beside close to the phase transitions \cite{notesystsize}. Due to the shape of $\langle K_\perp \rangle$, one can find non-trivial solutions above a certain value of the pump strength $A\approx 4.6J_\|/L$ (see Fig.~\ref{fig:kpersol09}(b)), where the cavity field takes a finite value. The coupling between a many-body system and an optical cavity can result in the presence of bifurcation points in its phase diagram~\cite{Tian2016, GuckenheimerHolmes1983}. In the case of $L=120$ we can see that around $A\approx 4.6J_\|/L$ the steady state diagram shows a transition from a trivial stable solution to one of the two nontrivial solutions, $J_\perp(A)$ or $-J_\perp(A)$. In the regions where the phase transitions take place $5.3J_\|/L\lesssim A\lesssim 5.4J_\|/L$ and $6.07J_\|/L\lesssim A\lesssim 6.1J_\|/L$, multiple solutions of the self-consistency condition can exist for the same value of the pump strength. Due to the limited numerical resolution, we cannot decide which solutions are stable. 

\begin{figure}[hbtp]
\centering
\includegraphics[width=.5\textwidth]{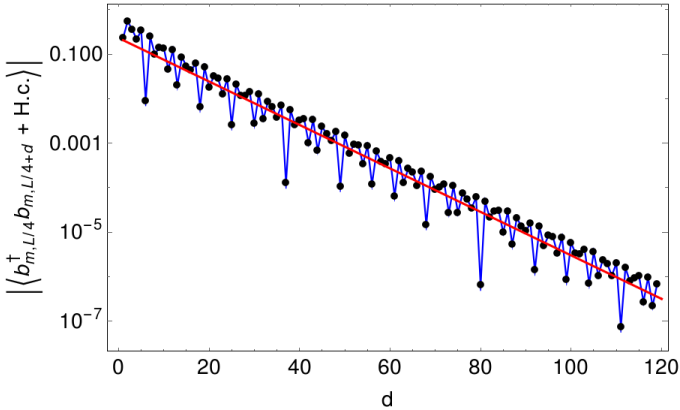}
\caption{The absolute value of the single particle correlations, $|\langle b^\dagger_{m,L/4} b_{m,L/4+d}+\text{H.c.} \rangle|$, in a semi-logarithmic plot, in the Meissner phase, for $L=240$, $A=6.7 J_\|/L$, $\varphi=0.9\pi$, $\rho=0.5$, $U=1.5J_\|$, and $V_{trap}=0$. The correlations decay exponentially, which signals the Mott-insulating phase. The red curve is the fit $\propto e^{-\alpha x}$, with the fit parameter is $\alpha=0.112\pm 0.02$. 
 }
\label{fig:corr09}
\end{figure}

In the following, we characterize the steady states which correspond to the stable solutions.
In Fig.~\ref{fig:observables09}(a), we can see that below $A\lesssim 6.1J_\|/L$ the rung currents take finite values which indicates a vortex state. A clear abrupt change of the rung current and other observables signals another transition between states at approximately $5.3J_\|/L\lesssim A\lesssim 5.4J_\|/L$. The vortex state in between $5.4J_\|/L \lesssim A\lesssim 6.1J_\|/L$ is characterized by a stable vortex density  $\rho_{v}=1/2$ which points towards a $\text{VL}_{1/2}\text{-SF}$ state. This state is confirmed by the value of the central charge $c\approx 1$ (Fig.~\ref{fig:observables09}(b)) and the algebraic decay of the single particle correlation functions (not shown). We observe that in the $\text{VL}_{1/2}\text{-SF}$ phase the chiral current changes its sign due to the increase of the unit cell, as explained in Refs.~\cite{GreschnerVekua2015, GreschnerVekua2016}. The incommensurate vortex density below  $A\lesssim 5.4J_\|/L$ suggests a vortex liquid state. It would have a central charge of two. However, the value of the central charge extracted from our numerical calculations still depends crucially on the system size as shown in Fig.~\ref{fig:observables09}(b) such that a final conclusion is difficult. 

For a pump strength above $A\gtrsim 6.1J_\|/L$ the rung current is suppressed and the current flows only along the legs of the ladder.
The central charge vanishes indicating a totally gapped system. This is confirmed by the exponential decay of the single-particle correlations with distance as seen in Fig.~\ref{fig:corr09}. This state is a Meissner Mott-insulator.

Thus for the parameters $\varphi=0.9\pi$, $\rho=0.4$ and $U=1.5J_\|$ the dynamical stabilization of a vortex liquid superfluid, vortex lattice superfluid with $\rho_v=1/2$ and a Meissner Mott-insulating states is possible.

\begin{figure}[hbtp]
\centering
\includegraphics[width=.5\textwidth]{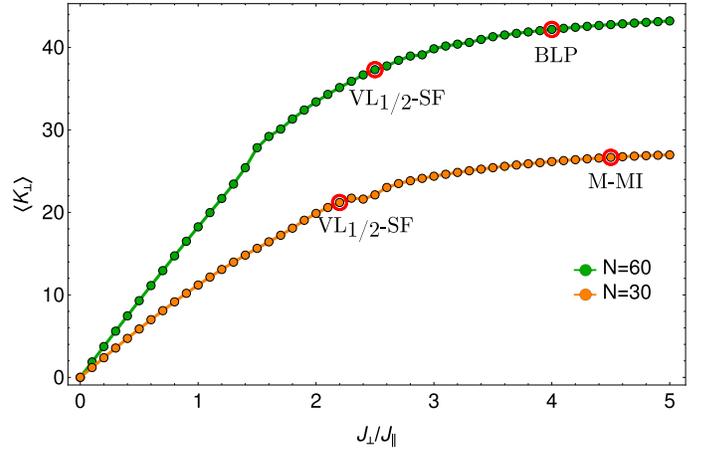}
\caption{The expectation value of the directed rung tunneling  $\langle K_\perp \rangle$ in the presence of a harmonic trapping potential, $V_{trap}=1.5J_\|$, for the parameters $\varphi=0.9\pi$, $U=1.5J_\|$ and $L=60$, for two particle numbers, $N= 60$ and $N= 30$. For the points marked with red circles we identified the phases present in the center part of the trap and they are discussed in the main text.
 }
\label{fig:kpertrap}
\end{figure}

\begin{figure}[hbtp]
\centering
\includegraphics[width=.5\textwidth]{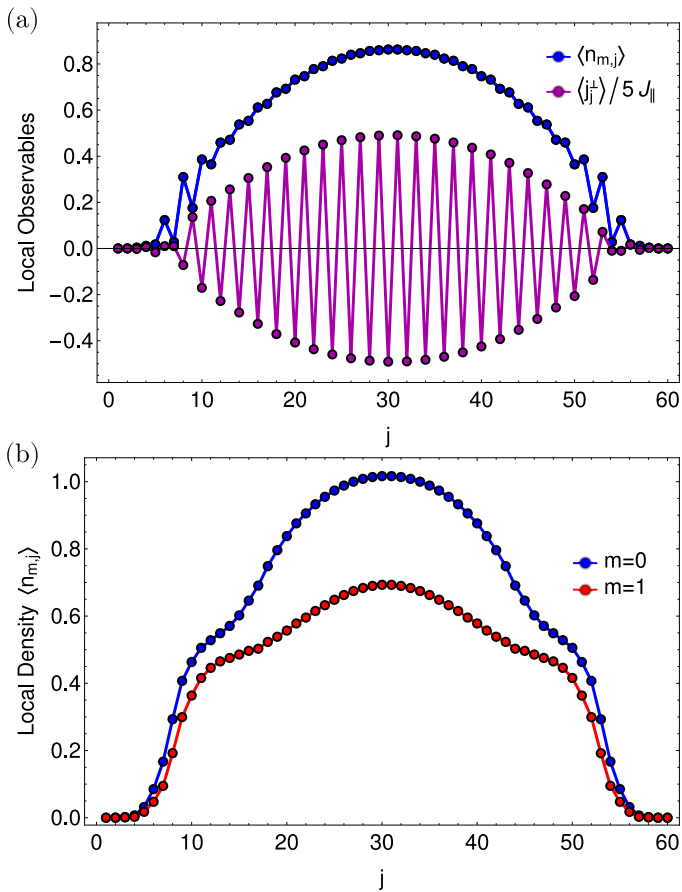}
\caption{(a) The expectation value of the local density $\langle n_{m,j} \rangle$ (blue line) and the local rung current $\langle j_{j}^\perp \rangle$ (purple line), for $J_\perp=2.5J_\|$, in the $\text{VL}_{1/2}\text{-SF}$ state. (b) The expectation value of the local density $\langle n_{m,j} \rangle$ on the two legs of the ladder $m=0$ (blue line) and $m=1$ (red line), for $J_\perp=4J_\|$, in the biased ladder phase.
The parameters used are $\varphi=0.9\pi$, $U=1.5J_\|$, $N=60$ and $L=60$, in the presence of a harmonic trapping potential, $V_{trap}=1.5J_\|$.
 }
\label{fig:trap1}
\end{figure}

\subsection{\label{sec:trap}The effect of a harmonic potential on the steady states}

In the previous sections we have focused on the homogeneous case without any trapping potentials, in this subsection we will consider the effect of a harmonic trapping potential $H_{trap}$ on the steady states. 

\begin{figure}[hbtp]
\centering
\includegraphics[width=.5\textwidth]{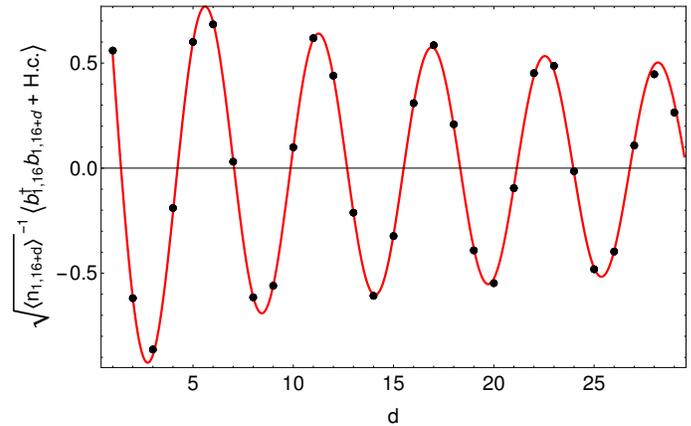}
\caption{The value of the single particle correlations, weighted by the inhomogeneous density, $\sqrt{\langle n_{1,16+d}\rangle}^{-1}\langle b^\dagger_{1,16} b_{1,16+d}+\text{H.c.} \rangle$, in the biased ladder state. The correlations decay algebraically, which signals the superfluid state. The red curve is the fit $\propto x^{-\alpha}\cos(\beta x)$, with the fit parameters $\alpha=0.263\pm 0.007$ and $\beta=1.114\pm 0.001$. The parameters used are $J_\perp=4J_\|$, $\varphi=0.9\pi$, $U=1.5J_\|$, $N=60$ and $L=60$, in the presence of a harmonic trapping potential, $V_{trap}=1.5J_\|$ (the same as in Fig.\ref{fig:trap1}(b)).
 }
\label{fig:trapcorr1}
\end{figure}

\begin{figure}[hbtp]
\centering
\includegraphics[width=.5\textwidth]{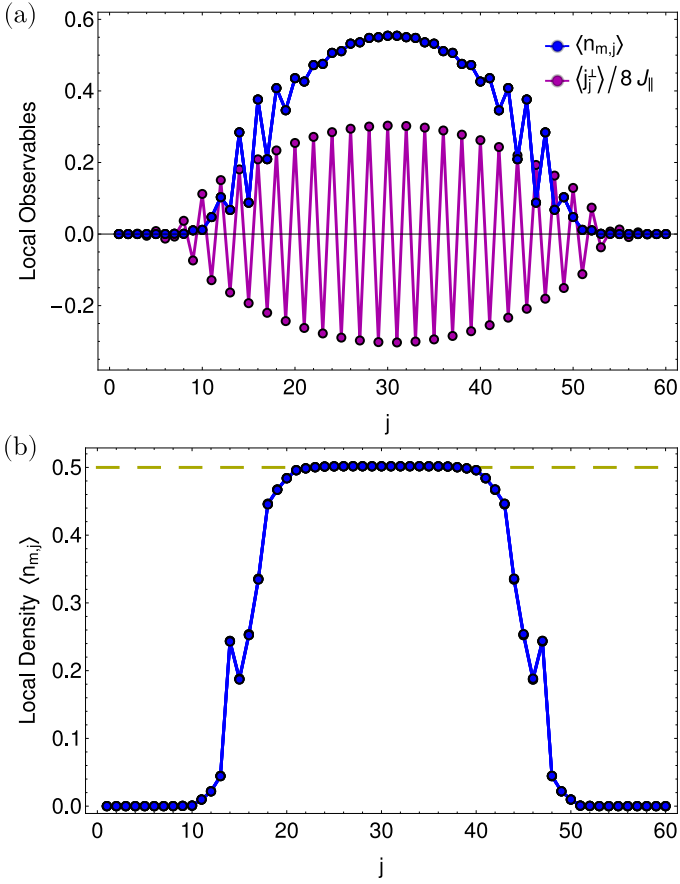}
\caption{a) The expectation value of the local density $\langle n_{m,j} \rangle$ (blue line) and the local rung current $\langle j_{j}^\perp \rangle$ (purple line), for $J_\perp=2.2J_\|$, in the $\text{VL}_{1/2}\text{-SF}$ state. (b) The expectation value of the local density $\langle n_{m,j} \rangle$, for $J_\perp=4.5J_\|$, in the Meissner-Mott insulating state. The parameters used are $\varphi=0.9\pi$, $U=1.5J_\|$, $N=30$ and $L=60$, in the presence of a harmonic trapping potential, $V_{trap}=1.5J_\|$. The dashed horizontal line indicates the constant value $0.5$.
 }
\label{fig:trap2}
\end{figure}

\begin{figure}[hbtp]
\centering
\includegraphics[width=.5\textwidth]{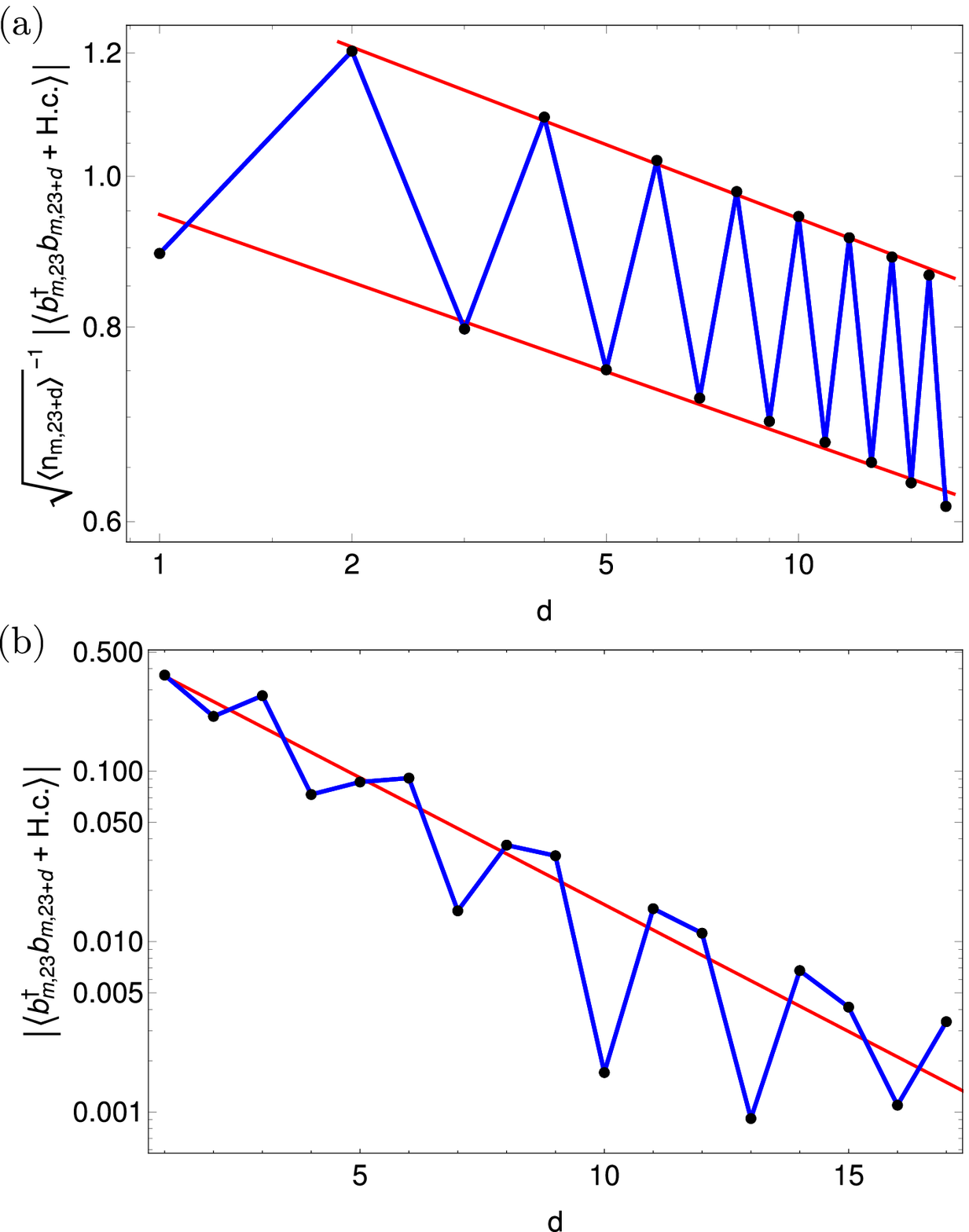}
\caption{(a) The absolute value of the single particle correlations, weighted by the inhomogeneous density, $\sqrt{\langle n_{m,23+d}\rangle}^{-1}|\langle b^\dagger_{m,23} b_{m,23+d}+\text{H.c.} \rangle|$, in a logarithmic plot, for $J_\perp=2.2J_\|$. The correlations decay algebraically, which signals the superfluid state. The red curve is the fit $\propto x^{-\alpha}$, with the fit parameter $\alpha=0.15\pm 0.01$.
(b) The absolute value of the single particle correlations, in a semi-logarithmic plot, for $J_\perp=4.5J_\|$. The correlations decay exponentially, which signals the Mott-insulating state. The red curve is the fit $\propto e^{-\alpha x}$, with the fit parameter $\alpha=0.34\pm 0.04$. 
The parameters used are $\varphi=0.9\pi$, $U=1.5J_\|$, $N=30$ and $L=60$, in the presence of a harmonic trapping potential, $V_{trap}=1.5J_\|$. 
 }
\label{fig:trapcorr2}
\end{figure}

In the following, we will discuss which steady states can be dynamically organized for flux ${\varphi=0.9\pi}$, on-site interaction $U=1.5J_\|$ and two values of the total particle number $N= 60$ and $N=30$, in a harmonic trap of strength $V_{\text{trap}}=1.5J_\|$. Due to the trapping potential, the density of the atoms varies throughout the system and a coexistence of different states can be found across the trapped region. Thus, in order to identify the states we analyze local observables, as the local density or the local rung currents, Eq.~(\ref{eq:localcur}). We focus on the states that are present in the center region of the trap, where the gradient of the trapping potential is relatively small. The simulated system size has been chosen such that the density decays to zero before reaching the boundary.

In Fig.~\ref{fig:kpertrap} the expectation value of the directed tunneling $\langle K_\perp \rangle$ has been plotted for the two values of the filling. We can see that we have stable steady states for $J \gtrsim 1.5 J_\|$ (for $N=60$) and  $J \gtrsim 1J_\|$ (for $N=30$). We marked in Fig.~\ref{fig:kpertrap} the points for which we identify the nature of these stable steady states. In Fig.~\ref{fig:trap1}(a) the density profile along one leg of the ladder and the local rung current pattern are depicted for the case with particle number $N=60$, which leads to a central filling of $\rho\approx0.86$ and $J_\perp=2.5J_\|$, corresponding to a pump strength of $A\approx0.067J_\|$. One can observe a varying density distribution along the leg of the ladder, which points towards a compressible state, namely a superfluid phase. The rung current pattern in the central region exhibits commensurate vortices with a vortex density of $\rho_v=1/2$. Thus, we identify this state as the $\text{VL}_{1/2}\text{-SF}$. For the same particle number $N= 60$ and $J_\perp=4J_\|$, corresponding to pump strength $A\approx0.094J_\|$, the density profile for each leg of the ladder has been plotted in Fig.~\ref{fig:trap1}(b). We observe that a density imbalance between the two legs of the ladder is present throughout the ladder, except at the boundaries where the density is close to $0$. The finite density imbalance indicates that in the central region a biased ladder state forms. The superfluid nature of this state is confirmed by the algebraic decay of the single particle correlation function, weighted by the inhomogeneous density \cite{KollathZwerger2004}, along one of the legs of the ladder (see Fig.~\ref{fig:trapcorr1}). 

In the case of $N=30$ and $J_\perp=2.2J_\|$, corresponding to a pump strength of $A\approx0.103J_\|$, we also identify a $\text{VL}_{1/2}\text{-SF}$ state in the center of the trap, by analyzing the local density profile and the rung current configuration depicted in Fig.~\ref{fig:trap2}(a). The single particle correlation function, weighted by the inhomogenous density, decays algebraically in this superfluid state. By increasing the pump strength to $A\approx0.168J_\| $ with self-consistent solution of $J_\perp=4.5J_\|$, an incompressible state is formed in the center of the trap which is indicated by the density plateau at filling $\rho=0.5$ as shown in Fig.~\ref{fig:trap2}(b). The exponential decay of the single particle correlations in the plateau region (see Fig.~\ref{fig:trapcorr2}(b)) and suppressed rung currents, confirm the presence of a Meissner Mott-insulator in the center of the ladder which is stabilized via the coupling of the atoms to the cavity. The Mott-insulating state is surrounded by a small region of a superfluid.

Thus, we confirm that even in the presence of a harmonic trapping potential the dynamical stabilization of vortex lattice superfluid with $\rho_v=1/2$, biased ladder and Meissner Mott-insulating states is achievable. We expect that other interesting states are similarly robust and can also be stabilized in the presence of a harmonic trapping potential. 

\section{\label{sec:conclusion}Conclusions}

In this work we investigated the steady state diagram of bosonic atoms coupled to an optical cavity. The bosonic atoms are confined to quasi-one-dimensional ladder structures with a large potential offset between the legs formed by optical lattices in the presence and abscence of a harmonic trapping potential. Due to the chosen coupling of the atoms and the cavity field, the tunneling on the rungs takes place via the generation or annihilation of a cavity photon and has a spatial dependent phase imprint, which breaks the time reversal symmetry. Above a critical value of the pump strength, due to the feedback between the atoms and the cavity field, a finite occupation of the cavity field mode arises and the bosonic atoms feel an artificial gauge field. We demonstrated the stabilization of Meissner phases both in the superfluid (M-SF) and in the Mott insulator (M-MI) regimes.
Additionally, we find vortex superfluid phases which can be incommensurate (V-SF), or commensurate with the ladder ($\text{VL}_{\rho_v}\text{-SF}$), the vortex lattice states. 
Finally, a biased-ladder superfluid phase with imbalanced density on the two legs of the ladder is also stabilized. 
Furthermore, we show that in the presence of the harmonic trapping potential, a vortex lattice superfluid $\text{VL}_{\rho_v}\text{-SF}$, a biased ladder and a Meissner Mott-insulating states are stablized in the center of the trap. 

One of the advantages of the dynamic stabilization of the states is their robustness. The evolution towards the steady states is characterized by an dissipative attractor dynamics which means that many external perturbations will decay exponentially towards the steady state. The extension of the presented scheme to a two-dimensional geometry is of interest since the ground state of the two-dimensional Bose-Hubbard model in an artificial magnetic field can exhibit exciting phases like vortex lattice superfluid phases with different vortex configurations, that are breaking the spatial symmetry, \cite{GoldbaumMueller2008, GoldbaumMueller2009, PowellSarma2011}, or bosonic integer and fractional quantum Hall states \cite{HeVishwanath2017}. 

Experimentally, the proposed setup can be implemented in different ways. We mentioned the ladder structures created using superlattice potentials, however, an alternative method is to use the synthetic lattice dimension. In this implementation the atoms are confined to one-dimensional structures and the second direction is implemented by internal states of the atoms \cite{CeliLewenstein2014, ManciniFallani2015, StuhlSpielman2015}. The transitions between the two states are implemented via Raman transitions employing the cavity mode and an external pump beam.

In order to distinguish the different steady states one can perform different measurements.  A very important measurement is the occupation of the cavity field which can be obtained from the leaking of the photons of the cavity. This already gives the distinction between the trivial and non-trivial self-organized phases and identifies by this the dynamically organized gauge field. 
Additionally, one would like to obtain more information on the state of the atomic component. There are different possibilities and depending on the experimental setup and the state one is particularly interested in. From insitu measurements of local densities, one can identify the biased ladder phase and obtain information on the different vortex patterns. In Ref.~\cite{AtalaBloch2014}, Atala et al. (2014) measure the chiral currents and the momentum distribution, from a measurement schedule involving the projection onto double wells. Furthermore, in Ref.~\cite{GreschnerVekua2016} by Greschner et al. (2016) a connection between the peak in the momentum distribution and the vortex density is drawn. From the vortex density one can distinguish the vortex liquids from the vortex lattices. 
Additionally to these destructive measurement of the the atomic component, a non-destructive method to measure the chiral current from the photons leaking from the cavity has been introduced \cite{KollathBrennecke2016}. This uses a probe beam and another cavity mode in which the tunneling along chain of the ladder is coupled to the empty probe cavity mode and the chiral current can, thus, be directly measured by observing the appropriate quadrature using a heterodyne detection scheme. 

\section*{\label{sec:acknowledgments}Acknowledgments}
We thank F. Heidrich-Meisner, U. Schollw\"ock, and S. Wolff 
for fruitful discussions. 
We acknowledge financial support from the DFG (individual research grant, FOR 1807 and TR 185).


\begin{thebibliography}{10}
\bibitem{notesystsize}
For smaller systems and $J_\perp<1.8$ it is not clear that a stable solution exists for all values.
\end{thebibliography}
\end{document}